\documentclass{elsart}

\usepackage{graphicx}

\usepackage{amssymb}

\begin{document}

\begin{frontmatter}

\title{Nuclear mass systematics by complementing the Finite Range Droplet Model with neural networks}

\author{S.~Athanassopoulos},
\author{E.~Mavrommatis}
\address{Physics Department, Division of Nuclear \& Particle Physics, University of Athens, GR--15771 Athens, Greece}
\author{K.~A.~Gernoth}
\address{School of Physics \& Astronomy, University of Manchester, M13 9PL, UK}
\author{J.~W.~Clark}
\address{McDonnell Center for the Space Sciences and Department of Physics, Washington University, St.\ Louis, Missouri 63130, USA}

\begin{abstract}
A neural-network model is developed to reproduce the differences
between experimental nuclear mass-excess values and the 
theoretical values given by the Finite Range Droplet Model. 
The results point to the existence of subtle regularities of 
nuclear structure not yet contained in the best 
microscopic/phenomenological models of atomic masses.  
Combining the FRDM and the neural-network model, we 
create a hybrid model with improved predictive performance 
on nuclear-mass systematics and related quantities.
\end{abstract}

\end{frontmatter}

\def\sigmarms{$\sigma_{\mathrm rms}$}

\section{Introduction}
The problem of devising global models of nuclidic (atomic) masses (see Ref.~\cite{Lunney} for a recent review) is of great current interest in connection with experimental studies of nuclei far from stability conducted at heavy-ion and radioactive ion-beam facilities and with the theory of nucleosynthesis and supernova explosions \cite{opp}.  The spectrum of global mass models ranges from those with high theoretical input that explicitly take account of known physical principles in terms of a relatively small number of fitting parameters, to models that are shaped only by the 
data and thus have a correspondingly large number of adjustable parameters. Current models of the former class that 
define the state of the art are the Finite Range Droplet Model (FRDM) of M\"oller, Nix, and coworkers \cite{FRDM1} and the Hartree-Fock-Bogoliubov model (HFB2) of Pearson, Tondeur and coworkers \cite{HFB2}.  Statistical models based on neural networks are situated far toward the other end of the spectrum.  They have been under continuing 
development in recent years, to the extent that they can now provide a valuable 
complement to conventional global models \cite{stat}.

Here we provide a preliminary report of results from a synthesis \cite{new}  
of the two approaches.  Training by example, a neural network is constructed
that estimates the differences $\Delta M^{\rm exp}- \Delta M^{\rm FRDM}$ 
between experiment and the FRDM, where $\Delta M$ denotes the nuclidic mass excess. 
Combining the FRDM with this neural network, we obtain a hybrid global
mass model that performs with precision both in reproducing $\Delta M$ values
for familiar nuclei and predicting them for new nuclei.
This strategy is pursued with the hope of determining whether the 
residual physical corrections to the FRDM model (a) stem from a
large number of small effects that may fluctuate strongly with $Z$
and $N$, defying systematic quantification, or instead (b) can be 
attributed in part to regularities of nuclear structure not yet 
embodied in theory. 

\section{Neural-network model of the mass differences}
\begin{figure*}[b]
\begin{center}
\includegraphics[scale=.8]{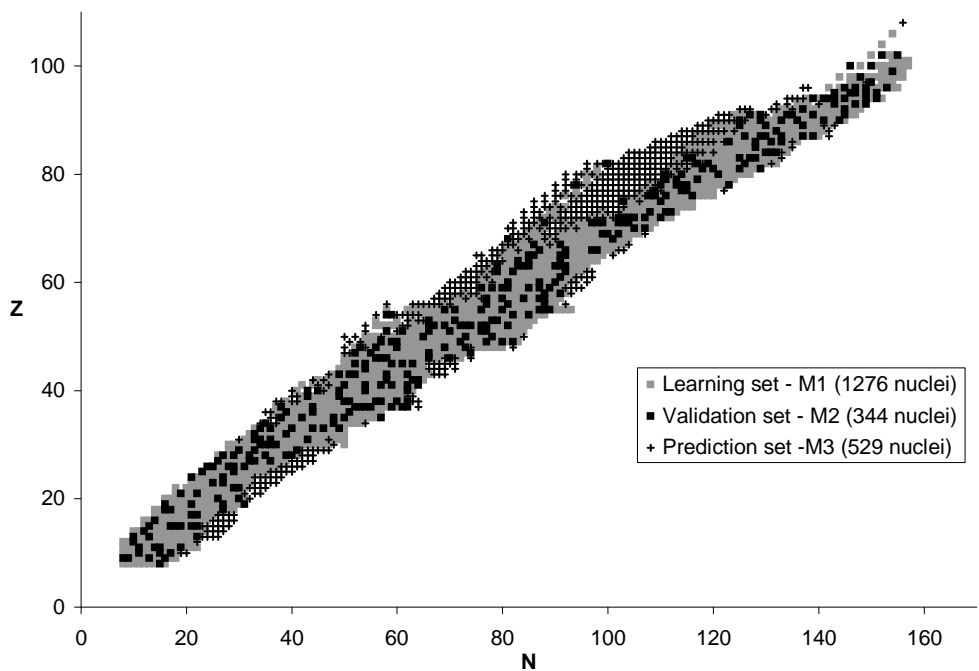}
\end{center}
\caption{\label{fig:figure1}Locations in the $N-Z$ plane are indicated for the M1, M2, and 
M3 data sets employed in neural-network modeling of the differences 
between experimental mass-excess values and those given by the FRDM.}
\end{figure*}
A multilayer feedforward architecture is adopted for the neural network, 
having the structure indicated schematically by ($4$--$6$--$6$--$6$--$1$)[169].  
The four input units encode the atomic number $Z$, the neutron number $N$, and their 
respective parities.  The single output unit encodes the mass-excess 
difference $\Delta M^{\rm exp}-\Delta M^{\rm FRDM}$.  Three intermediate
layers, each containing six units, transfer information from input
to output through weighted connections.  The total number of 
weight parameters charactering the network is 169.
To construct the neural-network model, we have employed the database of 
1654 nuclei fitted by the FRDM parameterization of Ref.~\cite{FRDM1}, 
screening out some uncertain cases in light of the more recent 
experimental mass-excess assignments published in the 2003 Atomic Mass Evaluation (AME03) \cite{Ame03}. 
The surviving $1620$ nuclei are divided randomly into two data sets 
of $1276$ (M1) and $344$ (M2) nuclei, which respectively comprise the
learning and validation sets for neural-network modeling.  Performance
on the learning set serves as the criterion for progressive adjustment
of the weights of the feedforward connections, while performance 
on the validation set is used to guide the termination of training.  To 
obtain an unambiguous measure of predictive performance, some of 
the data must be reserved as a test set, or prediction set, 
which is never referred to during the training process.  The test
set (denoted M3) is provided by the remaining $529$ nuclei 
of the AME03 evaluation.  These data points correspond predominantly
to nuclides far from stability, lying on the outer fringes of the $1620$-nuclide 
set $M1 \cup M2$  as viewed in the $N-Z$ plane (see Fig.~1).  The ability 
of the neural network to model the difference $\Delta M^{\rm exp}-\Delta M^{\rm FRDM}$ 
is illustrated in Fig.~2.
\begin{figure*}[b]
\begin{center}
\includegraphics[scale=.8]{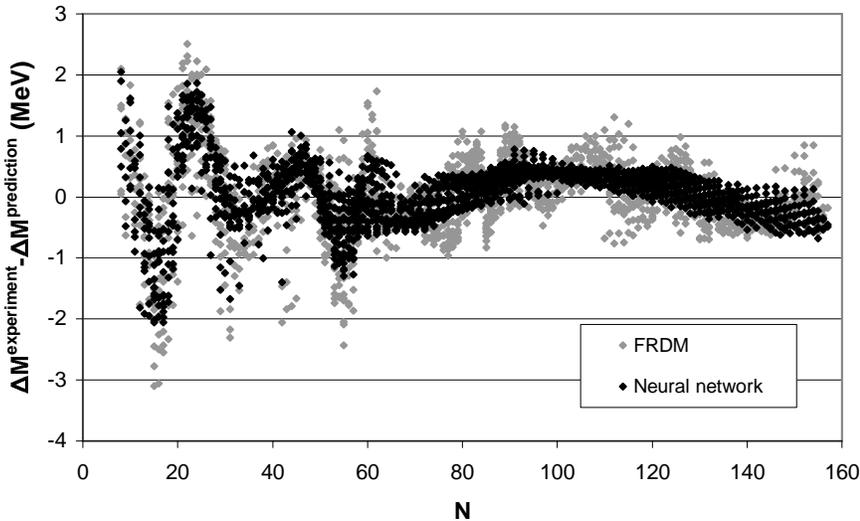}
\end{center}
\caption{\label{fig:figure2}Mass-excess differences between experiment and FRDM for the 
data sets M1 and M2 involved in the training process are compared with the corresponding differences predicted by the neural network.}
\end{figure*}

It is seen that the deviations of the FRDM evaluation from experiment for the data sets M1 and M2 involved 
in the training process can be substantially reproduced by the neural
network.
\vspace{-10truept}
\section{Mass excess evaluation -- Hybrid Model}
To generate and predict mass-excess values for nuclides of specified $Z$ and $N$, we construct a 
hybrid model by combining the FRDM outputs with the difference values predicted by 
the neural-network model described in Section 2.  In Table 1 we compare performance 
(measured by the rms error $\sigma_{\rm rms}$) on the learning, validation, 
and prediction sets for (i) the hybrid model, (ii) the neural-network mass 
model of Ref.~\cite{stat} and its most recent version \cite{new}
and (iii) the theoretical models FRDM \cite{FRDM1} and HFB2 \cite{HFB2}. The 
neural-network mass models of Refs.~\cite{stat,new} were trained to directly 
predict mass-excess values (as opposed to differences of mass-excess values).  
Likewise, the FRDM and HFB2 models were fitted to mass-excess data.

Overall, the hybrid model shows the best performance among the four
models considered, having very small error figures even for the 
prediction set M3. Further insight into the behavior of the hybrid
model of mass excess is furnished by Fig.~3, where the rms
error of the difference estimate {\it per isotope chain}, i.e.,
calculated for all $N$ for given $Z$, is plotted for the full database $M1 \cup M2 \cup M3$. For the majority of 
chains, the hybrid model yields smaller errors than FRDM and HFB2.
\begin{table}[b]
\caption{Root-mean-square error \sigmarms(MeV) in estimation of
mass excess by global models (see text for details).}
\begin{tabular}{cccc} 
\hline
Model & 
\shortstack{Learning set   \\(M1)} & 
\shortstack{Validation set \\(M2)} &
\shortstack{Prediction set \\(M3)} \\ 
\hline
FRDM (\cite{FRDM1})		     & 0.68 & 0.71 & 0.58 \\ 	
HFB2 (\cite{HFB2})		     & 0.67	& 0.68 & 0.67 \\
Neural net mass model (\cite{stat})& 0.44	& 0.44 & 0.95 \\
Neural net mass model (\cite{new}) & 0.28	& 0.40 & 0.71 \\
Hybrid model 			     & 0.40	& 0.49 & 0.41 \\
\hline
\end{tabular}
\end{table}
\begin{figure*}[t]
\begin{center}
\includegraphics[scale=.8]{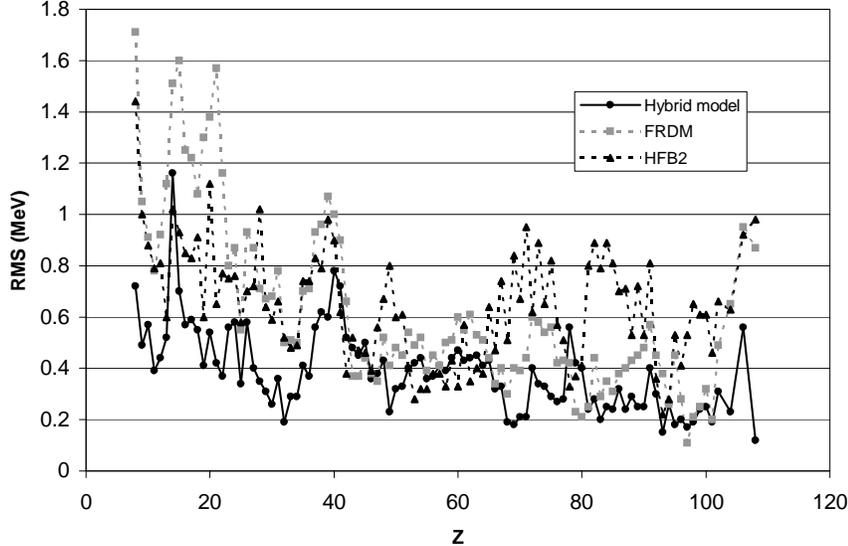}
\end{center}
\caption{\label{fig:figure1}RMS mass-excess error per isotope chain, plotted versus atomic number 
$Z$ for the full AME03 database.  Results are shown for the hybrid,
FRDM, and HFB2 global mass formulas.}
\end{figure*}
\section{Mass-related nuclear quantities -- Hybrid Model}
Mass-related quantities of interest can also be evaluated based
on the various models of mass excess, statistical and theoretical. Table 2 
presents the rms errors in determination of the one- and two-proton 
separation energies $S(p)$ and $S(2p)$, the one- and two-neutron separation 
energies $S(n)$ and $S(2n)$, and the $Q$-values for alpha and beta-minus 
decays, for all nuclei in AME03 with experimentally measured values.
The hybrid model outperforms its competitors in all of the comparisons,
although generally by smaller margins than for the mass excess (cf.\
Table 1). However, the ultimate test of any global model is in the
accuracy it can achieve on nuclei that have not been used in 
adjusting its parameters.  Table 3 reports rms errors in the separation
energies and $Q$-values for the subset of cases involving {\it only}
nuclides of the prediction set $M3$.  In this part of the nuclidic
chart, the hybrid model demonstrates predictive performance comparable 
to that of FRDM alone.  

\begin{table}[b]
\caption{Performance of global mass models for various quantities 
related to nuclear-mass systematics, quantified by the corresponding rms 
error over all cases involving AME03 nuclides \cite{Ame03} for which 
experimentally measured values are available.  Numerical entries are in MeV.}
\begin{tabular}{ccccccc} 
\hline
Model & 
\shortstack{$S(p)$\\(1968)}& 
\shortstack{$S(2p)$\\(1836)}&
\shortstack{$S(n)$\\(1988)}&
\shortstack{$S(2n)$\\(1937)}&
\shortstack{$Q(\alpha)$\\(2039)}&
\shortstack{$Q(\beta^-)$\\(1868)}\\
\hline
FRDM (\cite{FRDM1})	     		& 0.40 & 0.49 & 0.40 & 0.51 & 0.61 & 0.50	\\ 	
HFB2 (\cite{HFB2})	     		& 0.49 & 0.51 & 0.47 & 0.46 & 0.55 & 0.60	\\
Neural net mass model (\cite{stat})	& 0.53 & 0.61 & 0.48 & 0.58 & 0.67 & 0.64	\\
Neural net mass model (\cite{new}) 	& 0.56 & 0.49 & 0.38 & 0.46 & 0.62 & 0.53	\\
Hybrid model  		     		& 0.36 & 0.40 & 0.35 & 0.42 & 0.48 & 0.42	\\
\hline
\end{tabular}
\end{table}
\section{Conclusions -- Future steps}
Global semi-empirical models of atomic masses have reached a stage
of sophisication such that sub-MeV accuracy is achievable in 
predicting the mass excess of newly created nuclides.  At 
this stage one is naturally led to inquire whether the
residual errors are ``chaotic'' or random in nature, arising 
from the fluctuation and interplay of a large number of 
small physical effects as well as some experimental error.  
We have addressed this question by creating a neural-network 
model that generates the difference between experimental 
mass excesses and the values given by a state-of-the-art global mass 
model, specifically, the Finite Range Droplet Model of M\"oller, Nix, and 
coworkers \cite{FRDM1}.  Our results suggest that a significant 
portion of the residual error (perhaps 30-40\%) can be treated systematically, 
i.e., some regularities remain to be extracted from the data.
More extensive neural-network studies aimed at revealing the 
statistical behavior of the discrepancy are needed to test 
this inference.   

The present work has shown that hybrid models built by supplementing
the FRDM evaluation with a trained neural network show promise
of accurate prediction of atomic masses far from stability,
as well as other nuclear properties required as input for
theories of nucleosynthesis and supernova explosions.
\begin{table}[h]
\caption{Performance of global mass models for various quantities 
related to nuclear-mass systematics, quantified by the corresponding rms 
error over all cases involving only nuclides of the prediction set M3.
Numerical entries are in MeV.}
\begin{tabular}{ccccccc} 
\hline
Model & 
\shortstack{$S(p)$\\(453)}& 
\shortstack{$S(2p)$\\(434)}&
\shortstack{$S(n)$\\(435)}&
\shortstack{$S(2n)$\\(418)}&
\shortstack{$Q(\alpha)$\\(465)}&
\shortstack{$Q(\beta^-)$\\(387)}\\
\hline
FRDM (\cite{FRDM1})	     		& 0.41 & 0.44 & 0.40 & 0.40 & 0.52 & 0.51	\\ 	
HFB2 (\cite{HFB2})	     		& 0.45 & 0.46 & 0.41 & 0.41 & 0.48 & 0.56	\\
Neural net mass model (\cite{stat})	& 0.65 & 0.70 & 0.70 & 0.78 & 0.48 & 0.50	\\
Neural net mass model (\cite{new}) 	& 0.65 & 0.65 & 0.54 & 0.67 & 0.48 & 0.50	\\
Hybrid model  		     		& 0.41 & 0.39 & 0.37 & 0.41 & 0.48 & 0.50	\\
\hline
\end{tabular}
\end{table}
\vspace{-5truept}
\begin{ack}
This research has been supported in part by the U.~S. National Science Foundation under Grant No.~PHY-0140316 and by the University of Athens under Grant No.~70/4/3309.  We thank J. Rayford Nix for suggesting the hybrid strategy
explored here.
\end{ack}

\end{document}